\documentclass[%
preprint,
amsmath,amssymb,
aps,
prb,
]{revtex4-2}

\usepackage{graphicx}
\usepackage{dcolumn}
\usepackage{bm}
\usepackage{ulem} 
\usepackage{url}
\usepackage[bottom]{footmisc}
\usepackage[caption=false,font=footnotesize]{subfig}
\usepackage{amsmath}
\usepackage{amssymb}
\usepackage{amsfonts}
\usepackage{breakcites}
\usepackage{xcolor}
\usepackage{tikz}
\usepackage[shortlabels]{enumitem}
\graphicspath{{./Figures/}}

\begin{document}

\title{A robust and efficient model for transmission of surface plasmon polaritons onto metal-insulator-metal apertures}

\author{S. B. \.{I}plik\c{c}io\u{g}lu}
\email{siplikcioglu15@ku.edu.tr}
\author{M. I. Aksun}
\email{iaksun@ku.edu.tr}
\affiliation{
	Department of Electrical and Electronics Engineering,\\
	Ko\c{c} University, Istanbul, Turkey
}

\begin{abstract}
A simple yet accurate model for the transmission of surface plasmon polaritons (SPPs) in a finite metal-insulator-metal (MIM) waveguide to the sides of the apertures is proposed and demonstrated to be more accurate than the available models. It is as simple as using a magnetic current density across the plane of the aperture whose value is defined by the SPPs with any number of modes in the waveguide through the equivalence principle. Then, the generated SPPs on both sides of the aperture are extracted from the convolution integral of the equivalent current density and Green’s function. As a result, the model provides the transmission coefficients of the SPPs in the MIM to the side walls of the aperture accurately and efficiently; not only for symmetric MIMs with a single isolating layer but also for non-symmetric ones with multi-layered insulating materials. The results are in very good agreement with those obtained by the FDTD method and better than the other approximations available in literature for a wide range of aperture widths.
\end{abstract}

\maketitle
\clearpage
\section{Introduction}
\begin{figure}[h]
	\centering
	{\includegraphics[width=16cm]{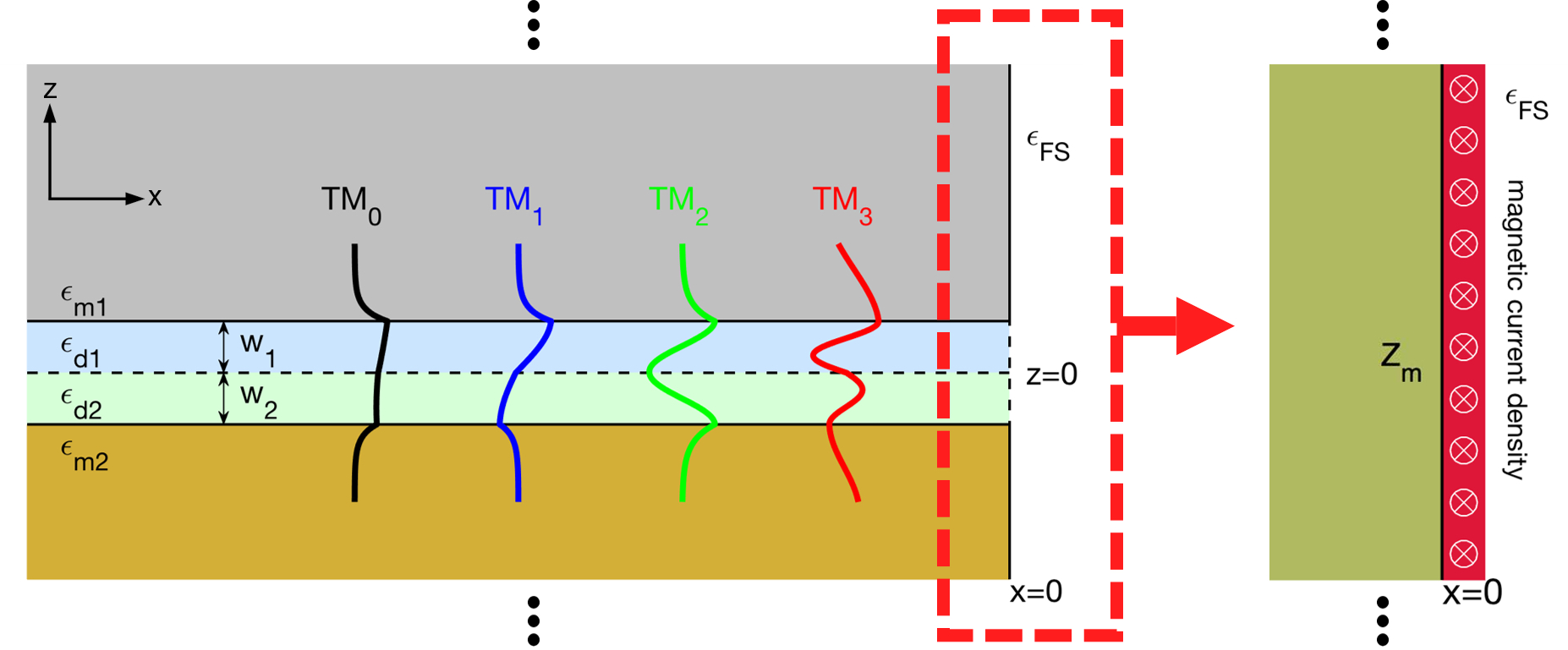}}
	\caption{A typical semi-infinite MIM waveguide with the SPP mode profiles and the effective surface impedance model with the impressed equivalent magnetic current density. The mode profiles are given at $\lambda_0=900$~nm for the silver-Si$\text{O}_2$-air-copper combination (from top to bottom) with the following material parameters: $\varepsilon_{m1}=-40.6923 + 0.7733i$, $\varepsilon_{m2}=-37.5145 + 0.5144i$, $\varepsilon_{d1}=2$, $\varepsilon_{d2}=1$, $w_1=w_2=100$ nm. }
	\label{fig:MIM_finite}
\end{figure}

Surface plasmon polaritons (SPPs) have been studied extensively in the past few decades, and deemed to have a great potential to help develop new applications/technologies in the fields of nanotechnology, communications and life sciences~\cite{Barnes2003,Ozbay2006,Stewart2008,Atwater2010}. Motivated by this outlook and the observations of extraordinary transmission through nano-holes in metals, there have been a flurry of studies on the transmission and reflection of SPPs at some canonical discontinuities, like holes, ridges, abrupt terminations or transitions to different waveguides or components~\cite{MartinMoreno2001,Oulton2007,Liu2008microscopic,Kocabas2008,GarciaVidal2010,Nikitin2010hole}. Although mostly rigorous full-wave methods, such as the finite-difference time-domain (FDTD) method, the Fourier-modal method, the mode matching method or their variants, have been used to characterize the discontinuities in these studies~\cite{Oulton2007,Liu2008microscopic,Kocabas2009MIM,Liu2010hybridwave,Gallinet2015}, some approximate models have also been proposed with success for efficient and yet accurate assessment of the structures~\cite{Zia2005,Veronis2005,Gordon2006,Lalanne2006approximate,Kocabas2008,Chandran2012reflection,Medina2008,Ozawa2019}. Inevitably, the approximate models are usually limited to either certain geometries or certain wavelength ranges. In this work, we propose an approximate model for nanoslits in metal-dielectric interface, modeled as metal-insulator-metal (MIM) waveguides opening up to a dielectric environment from both ends, as shown in Fig.~\ref{fig:MIM_finite}, to assess the transmission to the metal-dielectric surfaces beyond the edges of the aperture with better accuracy and applicability than the available models. 

As it is common in integrated optics, structures that are studied in the context of plasmonics generally involve discontinuities, either as a critical design choice or as an inevitable byproduct of the fabrication processes. Modeling these discontinuities poses a particular challenge due to the complexity of the problems involved, which generally have no analytical solutions. However, for plasmonics to gain importance in the design of integrated optics, its interaction with different geometrical constructs have to be analyzed, understood and modeled for practical use by the engineers. Towards this goal, there have been a plenty of work to characterize discontinuities in terms of reflection and transmission coefficients of the SPPs either on metal-insulator interface~\cite{Stegeman1981,Stegeman1983,Wallis1983,Jamid1997} or in metal-insulator-metal waveguides~\cite{Zia2005,Veronis2005,Gordon2006,Lalanne2006approximate,Kocabas2008,Chandran2012reflection}. Study of reflection and transmission of SPPs on a single metal-insulator interface goes back to the work in~\cite{Stegeman1981}, where the upper dielectric portion of the geometry is uniform along the interface while the lower metal portion consists of two pieces with different real dielectric constants as the discontinuity. As a continuation, a similar study on the SPP reflection for a uniform metal layer with two different dielectrics has also been reported~\cite{Stegeman1983}. These studies used the modal analysis and power matching at the discontinuity to extract the reflection and transmission coefficients, and a similar approach was later used to characterize the SPP reflection from a terminated lossless metallic half-plane~\cite{Wallis1983}. As a natural extension, the SPP reflectivity at step discontinuities was also studied using the method of lines~\cite{Jamid1997}. Following the discovery of extra-ordinary optical transmission~\cite{Ebbesen1998}, there have been a flurry of activities to describe the transmission through nano-holes and nano-slits in metal slabs~\cite{Porto1999,Takakura2001,Cao2002,Barnes2004,Chang2005}, which have provided ample computational tools and good intuitive understandings of the SPPs in the slits. Hence, these developments paved the way to model a nanoslit in metallic slab (considered as a MIM waveguide) to predict the reflection of SPPs in the slit and the transmission of SPPs from the slit to the SPPs on the nano-aperture sides (metal-dielectric interface)~\cite{Lalanne2005theory,Lalanne2006approximate}. As emphasized in~\cite{Lalanne2006approximate} and stated above for the modeling of discontinuities, it is of utmost importance to have an approximate and yet accurate and computationally efficient model for finite MIM structures, as is the main goal of the work presented in this paper.

To put things into perspective and to delineate the contributions of this work, it would be imperative to provide the contributions in~\cite{Lalanne2006approximate}, where the model for the reflection and transmission in a finite MIM waveguide was developed in two distinct intuitive steps: i. for the geometrical scattering part from the slit aperture, the metal layers are assumed to be perfect conductors as their dielectrics play less significant role in diffraction, and ii. once the near-field distribution is obtained, the transmission fields in the near vicinity of the slit aperture are obtained by using the mode orthogonality. Although this is a brilliant approach to get analytical expressions for the SPP generation, it is limited to a rather restricted class of slit geometries because of the approximate nature of the model, as also stated in~\cite{Lalanne2006approximate}. Restrictions stem from having to use symmetric geometries ($\epsilon_{m1} = \epsilon_{m2}$ in Fig.~\ref{fig:MIM_finite}), only the fundamental mode in the slit (black field profile in Fig.~\ref{fig:MIM_finite}) and homogeneous dielectric region in the slit ($\epsilon_{d1} = \epsilon_{d2}$ in Fig.~\ref{fig:MIM_finite}), all of which are addressed and resolved in the proposed approach. However, to be fair, the approximate approach presented in~\cite{Lalanne2006approximate} was specifically proposed to get analytic expressions for the transmission fields, while the main purpose of the work presented in this paper has been to achieve a computationally efficient and accurate model that is applicable to a rather broader spectrum of geometries, albeit at the expense of closed-form expressions at the end.

In this paper, assuming the reflection from the aperture of a MIM waveguide has been well approximated with the current vectorial mode-matching approach, an efficient, accurate and versatile numerical model for the transmission of the SPPs from the aperture onto the open metal-dielectric interfaces has been developed and compared with the full-wave method (FDTD) and the approximate model proposed in~\cite{Lalanne2006approximate}. The model is based on the equivalence principle with the impedance boundary and is able to employ as many modes in the MIM waveguide as needed with their full profiles. As a result, besides the accuracy, the model is robust, converges fast and can easily handle MIM structures with asymmetric metal configurations as well as planar dielectric layers as the insulating regions. 

\section{Theory}
For the sake of illustration, a general finite MIM configuration (or slit geometry) is shown in Fig.~\ref{fig:MIM_finite}, which captures almost all possible varieties of the geometries that have been studied in this work. The field profiles for the first four modes ($\text{TM}_0, \text{TM}_1, \text{TM}_2, \text{TM}_3$) have also been depicted on the figure as black, blue, green and red lines, respectively, as the higher order modes gain importance for asymmetric geometries as well as for symmetric ones with larger slit widths. Moreover, the equivalent magnetic current density and the model of the layered structure as a single effective medium is shown as the expanded view of the aperture plane ($x=0$) and its near vicinity, as it is the key component of the proposed model in this paper. 

The main idea of the proposed model, whose details will be presented in the following subsections, is based on the equivalence principle at the aperture plane. The radiation from the aperture due to the SPPs in the MIM waveguide is approximated by the radiation from an equivalent magnetic current density at the aperture plane backed by an effective impedance half-space in place of the MIM structure, as shown in Fig.~\ref{fig:MIM_finite}. Hence, the generated complex SPP amplitudes on both sides of the aperture ($z\ge w_1$ and $z\le -w_2$) can be easily obtained by analytically evaluating the surface wave contributions due to the magnetic current density at a reasonable distance from the aperture boundaries.

\subsection{Transmitted SPPs at the aperture plane}
\label{theory-transmit-spp}
Per Schelkunoff's surface equivalence principle, a radiation problem involving an aperture can be equivalently modeled for the region of interest (the right hand side of the aperture at $x=0$ plane in Fig.~\ref{fig:MIM_finite}) as a perfect electric conductor (PEC) on the side of the aperture that is not the region of interest, and an equivalent surface magnetic current density at the plane of the aperture~\cite{Rengarajan2000}. As a result, a rather complex geometry is transformed to a simpler geometry for which the fields in the region of interest can be obtained by a simple convolution integral of appropriate Green's function and the surface magnetic current density at the aperture plane. However, in the case of plasmonic apertures, replacing the impedance surface at the plane of the aperture with a PEC layer poses a problem, as PEC surface can not support the SPP modes, and as such, information pertaining to the near-field behavior of SPPs and quasi-cylindrical waves (QCW) would be lost. Instead, based on the modified equivalence formulation for an aperture on a flat impedance surface~\cite{Yoshitomi1994equivalent}, and its extension to apertures on general shapes~\cite{Glisson2002equivalent}, the equivalent magnetic surface current can be modified as follows:
\begin{equation}
	\textbf{M}^{eq}=\textbf{E}^{a} \times \hat{\textbf{n}} - ( Z_s \hat{\textbf{n}} \times \textbf{H}^{a}) \times \hat{\textbf{n}} 
	\label{equivalent-current-aperture}
\end{equation}
where $(\textbf{E}^{a},\textbf{H}^{a})$ and $Z_s$ are the fields at the aperture and the surface impedance of the metal cladding ($Z_s=Z_0/\sqrt{\varepsilon_m}$), respectively, and $\hat{\textbf{n}}$ shows the unit normal vector to the plane of aperture. Before moving forward, it would be necessary to restate the objective of this work for the sake of clarity of the discussions that will follow: the objective is to propose an approximate, accurate and efficient model for the transmission of SPPs from the aperture of a finite MIM waveguide to the side walls on both sides of the aperture. With reference to Fig.~\ref{fig:MIM_finite}, the goal is to find the transmission of SPPs in the MIM waveguide through the aperture onto the walls on both sides ($z > 0$ and $z < 0$) of the aperture at $x=0$ plane. Obviously, the impedance surface at the aperture plane may consist of two metal layers of same material (symmetric case) as well as of different materials (asymmetric case). Therefore, the equivalent surface impedance to be used in (\ref{equivalent-current-aperture}) for asymmetric cases can be approximated by the surface impedance of the metal at whose interface the transmission is to be calculated, that is,  $Z_s=Z_0/\sqrt{\varepsilon_{m1}}$ for the calculation of the transmission to the interface at $z>0$, and  $Z_s=Z_0/\sqrt{\varepsilon_{m2}}$ for the transmission to the interface at $z<0$. Consequently, keeping the discussion on the surface impedance in mind and using Fig.~\ref{fig:MIM_finite} as the reference, the equivalent magnetic surface current density at the aperture plane simplifies to
\begin{equation}
\textbf{M}_y^{eq}=\hat{\textbf{y}}\left[E^a_z - Z_s H^a_y\right]
\label{equivalent-current-tm}
\end{equation}
where $\hat{\bf n} = \hat{\bf x}$, $\textbf{E}^{a} =(E^a_x, E^a_z)$ and $\textbf{H}^{a} = \hat{\bf y} H^a_y$. Once the surface magnetic current density is defined, the total magnetic field can be deduced from the convolution integral of this source term with the associated component of the magnetic field Green's function as 
\begin{equation}
H_y^{tot}(x=0;z)= -\frac{\omega \varepsilon_0 \varepsilon_{FS}}{2 \pi } \int_{-\infty}^{\infty}  G^H_{yy} (x=0;z-z') M^{eq} (z') dz'
\label{equivalent-current-green}
\end{equation}
where the spatial-domain Green's function can be written as the integral of its spectral-domain counterpart, also known as the Sommerfeld integral~\cite{WCChewWaves,Gravel2008}:
\begin{equation}
G^H_{yy} (x=0;z) = \int_{-\infty}^{\infty} \frac{e^{i k_z z}}{2k_{xFS}}  \left[1+R^{TM}(k_z)\right] dk_z
\label{equivalent-current-green-cont}
\end{equation}
with $k_{xFS}=\sqrt{\varepsilon_{FS} k_0^2-k_z^2}$ and $R^{TM}$ being the Fresnel reflection coefficient for the metal-dielectric interface at $x=0$ plane. The evaluation of Sommerfeld integrals via numerical methods or analytical approximations has been thoroughly studied and documented, and with the advent of high-speed computers, they are most often evaluated numerically~\cite{Michalski2016Sommerfeld,AksunRobust,AksunAlparslan}. Nevertheless, the full-scale evaluation of the integral is not necessary as the present study only requires the surface wave pole contribution at the SPP pole of the integrand $(\beta_{MI}=k_0\sqrt{\frac{\varepsilon_{FS}\varepsilon_m}{\varepsilon_{FS}+\varepsilon_{m}}})$, for which the residue theorem provides the closed-form expression for the surface wave contribution of the Green's function~\cite{Nevels2014sommerfeld}:  
\begin{equation}
\begin{aligned}
G^H_{SPP} (x=0;z) = \left[ \frac{-2 \pi i \varepsilon_m}{\beta_{MI} \left( \frac{\varepsilon_m}{k^\beta_{xFS}} + \frac{\varepsilon_{FS}}{k^\beta_{xm}} \right)} \right] e^{i \beta_{MI} |z|} 
\end{aligned}
\label{equivalent-current-green-residue2}
\end{equation}
where $k^\beta_{xFS}=\sqrt{\varepsilon_{FS} k_0^2-\beta_{MI}^2}$ and $k^\beta_{xm}=\sqrt{\varepsilon_m k_0^2-\beta_{MI}^2}$. 
With the SPP contribution of the Green's function in closed-form, the transmitted SPP onto the extension of the aperture on both sides can be obtained by evaluating (\ref{equivalent-current-green}) at a reasonable distance away from the aperture edges. Note that this is the transmitted magnetic field due to the equivalent magnetic current density at the aperture plane, which in turn is due to the SPP in the MIM waveguide. Providing that the modal components in the MIM waveguide and the SPP field at the interface are normalized to have unit power, the resulting magnitudes of the magnetic field at $(x=0;z=\pm h)$ give the transmission coefficients referenced to $z = \pm h$, which could be re-referenced to the edges of the aperture by just de-embedding the distances between $\pm h$ and the edges at $z=w_1$ and $z= -w_2$. After having obtained the equivalent magnetic surface current density, which will  be detailed in the following subsection, the remaining integral (\ref{equivalent-current-green}) becomes quite trivial and can be evaluated analytically, not only because of the closed-form expression for the Green's function (\ref{equivalent-current-green-residue2}) but also due to the exponential nature of the SPPs along the transverse directions into the metals. 

\subsection{Equivalent magnetic current density}
One of the strengths of the equivalent current model comes from its adaptability to arbitrary input modes and mode-matching schemes: continuous or discretized waveguide modes can be used to compute the SPP transmission coefficient, as long as the approximate field expressions at the aperture are known. In this subsection, the generalized mode-matching approach for a layered MIM waveguide, as shown in Fig.~\ref{fig:MIM_finite}, is briefly introduced in order to obtain the field profile at the aperture and, in turn, to express the equivalent surface magnetic current density at the aperture plane.

Invoking the interface boundary conditions, total transverse electric and magnetic fields at the aperture can be written as a superposition of the incident fundamental mode and the reflected discrete and continuous spectrum of the waveguide modes as
\begin{equation}
	\begin{aligned}
	E^a_z=\bar{a}_0 E_z^{\text{TM}_0} + \sum_{n=0}^{\infty} a_{n} E_z^{\text{TM}_{n}} + \int_{0}^{\infty} a_\nu(k_z) E_z^{\nu} (k_z) dk_z \\
	H^a_y=\bar{a}_0 H_y^{\text{TM}_0} - \sum_{n=0}^{\infty} a_{n} H_y^{\text{TM}_{n}} - \int_{0}^{\infty} a_\nu(k_z) H_y^{\nu} (k_z) dk_z 
	\end{aligned}
	\label{mim_waveguide_modes2}
\end{equation}
where $a_{n}$ denotes the complex amplitude coefficient for the $n$-th order TM mode while $a_\nu$ denotes the complex amplitude function for the radiative modal continuum. While the radiative modes have been documented to influence the accuracy of the mode-matching method, particularly for more rigorous field-stitching schemes~\cite{Kocabas2009MIM,Oulton2007}, they introduce complexities to the approach proposed here that can not be justified by their contribution to the overall accuracy. Contrarily, it is shown here that a small number of guided modes would be sufficient to ensure convergence without compromising accuracy for a wide range of widths and wavelengths. Therefore, substituting the field approximations using a finite number of guided modes into (\ref{equivalent-current-aperture}) results in the equivalent magnetic current density given below:
\begin{equation}
{M}_y^{eq}= \left[\bar{a}_0 E_z^{\text{TM}_0} + \sum_{n=0}^{N-1} a_{n}E_z^{\text{TM}_{n}}\right] - Z_s \left[ \bar{a}_0 H_y^{\text{TM}_0} - \sum_{n=0}^{N-1} a_{n} H_y^{\text{TM}_{n}} \right]
\label{equivalent-current-aperture-cont}
\end{equation}
where $N$ is the number of modes used and their coefficients $a_{n}$'s are the unknowns to be determined. Using the continuity of the transverse electric fields at the aperture, the electric field components radiating beyond the aperture, evaluated at $x=0^+$, can be expressed by a superposition of the angular spectra of the individual modes: 
\begin{equation}
\begin{aligned}
E^a_z(x=0^+)=\frac{\bar{a}_0}{2\pi} \int_{-\infty}^{\infty} dk_z I_0(k_z) e^{ik_z z} + \sum_{n=0}^{N-1} \frac{a_{n}}{2\pi} \int_{-\infty}^{\infty} dk_z I_{n}(k_z) e^{ik_z z}  \\
\end{aligned}
\label{mim_waveguide_modes_diffract}
\end{equation}
where $I_{n}(k_z)=\int_{-\infty}^{\infty}E_z^{\text{TM}_{n}}e^{-ik_z z}dz$. Then, using the Maxwell's curl equation, an analogous expression for the transverse magnetic field can be obtained as
\begin{equation}
\begin{aligned}
H^a_y(x=0^+)=& \frac{-k_0^2}{2\pi\omega\mu_0} \bar{a}_0\int_{-\infty}^{\infty} dk_z \frac{I_0(k_z)\varepsilon_{FS}}{\sqrt{\varepsilon_{FS}k^2_0-k_z^2}} e^{ik_z z} +\\
 &\frac{-k_0^2}{2\pi\omega\mu_0} \sum_{n=0}^{N-1} a_{n} \int_{-\infty}^{\infty} dk_z \frac{I_{n}(k_z)\varepsilon_{FS}}{\sqrt{\varepsilon_{FS}k^2_0-k_z^2}} e^{ik_z z}  \\
\end{aligned}
\label{mim_waveguide_modes_diffract2}
\end{equation}
which leads to a set of algebraic equations with constant unknowns, 
\begin{equation}
\begin{aligned}
\sum_{n=0}^{N-1} a_{n}& \left[H_y^{\text{TM}_{n}}-\frac{k_0^2}{2\pi\omega\mu_0} \int_{-\infty}^{\infty} dk_z \frac{I_{n}(k_z)e^{ik_z z}\varepsilon_{FS}}{\sqrt{\varepsilon_{FS}k^2_0-k_z^2}}  \right] \\ =\bar{a}_0 & \left[ H_y^{\text{TM}_0}+\frac{k_0^2}{2\pi\omega\mu_0} \int_{-\infty}^{\infty} dk_z \frac{I_0(k_z)e^{ik_z z}\varepsilon_{FS}}{\sqrt{\varepsilon_{FS}k^2_0-k_z^2}} \right] \\
\end{aligned}
\label{mim_waveguide_modes_diffract3}
\end{equation}
This is a set of familiar equations where the method of moments (MoM) can be employed with ease to solve for the unknown coefficients~\cite{HarringtonMoM}. Although one of the simplest implementation of the MoM is via using the point-matching approach where the equality in (\ref{mim_waveguide_modes_diffract3}) is enforced over a set of discrete points, it usually converges slow and requires much more points than the number of unknowns that inevitably results in least-square implementation. Instead, one can use better (more natural for the problem in hand) testing functions that would improve convergence as well as simplify the computations. Therefore, the transverse electric field components of the modes would be better suited for the testing functions for the expression in (\ref{mim_waveguide_modes_diffract3}), where the transverse magnetic fields have already been used as the basis functions.  Hence, the resulting set of equations to be solved for unknown coefficients can be written as
\begin{equation}
\begin{aligned}
\sum_{n=0}^{N-1} a_{n}^* \left[ \langle E_z^{\text{TM}_{m}} , H_y^{\text{TM}_{n}}\rangle -\frac{k_0^2}{2\pi\omega\mu_0} \biggl \langle E_z^{\text{TM}_{m}} , \int_{-\infty}^{\infty} dk_z \frac{I_{n}(k_z)e^{ik_z z}\varepsilon_{FS}}{\sqrt{\varepsilon_{FS}k^2_0-k_z^2}} \biggr \rangle \right] \\
=\bar{a}_0^* \left[ \bigl \langle E_z^{\text{TM}_{m}}, H_y^{\text{TM}_0} \bigr \rangle +\frac{k_0^2}{2\pi\omega\mu_0} \biggl \langle E_z^{\text{TM}_{m}}, \int_{-\infty}^{\infty} dk_z \frac{I_0(k_z)e^{ik_z z}\varepsilon_{FS}}{\sqrt{\varepsilon_{FS}k^2_0-k_z^2}} \biggr \rangle \right]\\
\end{aligned}
\end{equation} 
where $\langle u,v \rangle=\int_{-\infty}^{\infty} u v^* dz$ and $m=0,1,2,...,N-1$. Note that the approach used in~\cite{Gordon2006,Lalanne2006approximate, Chandran2012reflection} is a special case of this formulation for a single mode.

As a final note on the theoretical part, all these formulations would make sense provided the guided modes of the structures studied in this work could be obtained with good accuracy, for which there have already been a number of numerical approaches, such as, Muller's method~\cite{BurdenNumerical}, argument-principle method~\cite{Kocabas2009MIM} and fixed-point interation~\cite{Kekatpure2009dispersion}, among others~\cite{Anemogiannis1999}.

\section{Results and discussion}
Before presenting the results, it would be instructive to set the stage for the following discussions by providing the details of the building blocks of the geometries, as shown in Fig.~\ref{fig:MIM_finite}. First off, throughout this work, the results obtained from the FDTD simulations by Lumerical's software have been used as the reference and depicted in all figures to assess the accuracy of the models employed~\cite{Lumerical}. Since the main building blocks of the geometries studied here are the metals, such as silver and copper, their dielectric properties have been drawn from the Drude model with the relevant material parameters (background permittivity $\varepsilon_{\infty}$, plasma frequency $\omega_p$ and collision frequency $\nu$) obtained through a simplex fit on the experimental data by Babar and Weaver~\cite{Babar2015}: $\varepsilon_{\infty}=3.965$, $\omega_p=9.2074$ eV and $\nu=0.0239$ eV for Ag, and $\varepsilon_{\infty}=9.1734$, $\omega_p=9.1235$ eV and $\nu=0.0265$ eV for Cu. 

For the sake of clarity of the presentation and completeness of the structural variants, the geometries that have been studied here have been divided into two main categories: symmetric and asymmetric finite MIM configurations, as the asymmetry may refer to different metal layers and/or to two different intermediate dielectric stacks (as on Fig~\ref{fig:MIM_finite}). Moreover, with no loss of generality, the space where the aperture opens up was chosen to be the free-space ($\varepsilon_{FS}=1$) for all the studies presented here.

\subsection{Symmetric MIM aperture}
\begin{figure}[h]
	\centering
	{\includegraphics[width=16.5cm]{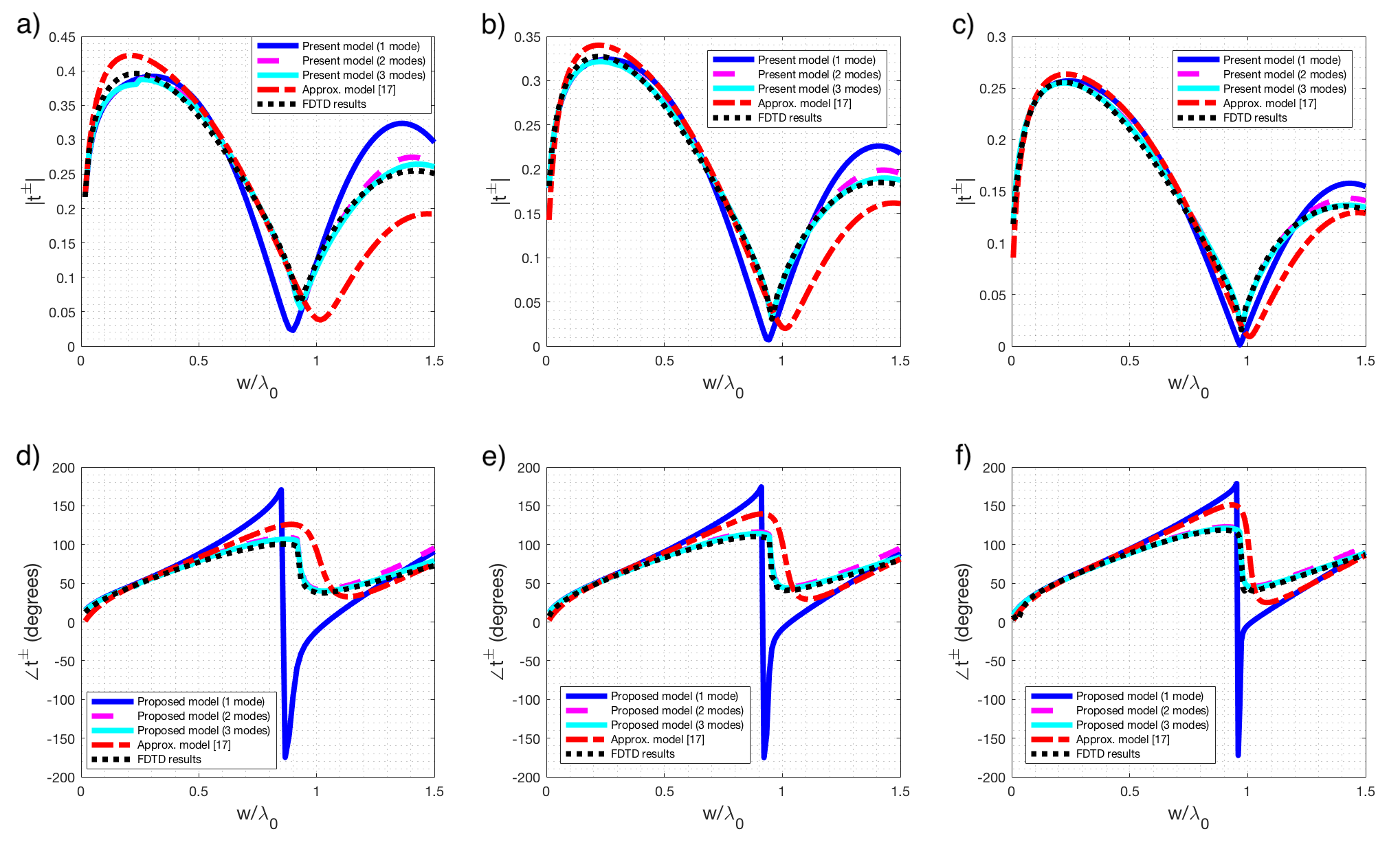}}
	\caption{Transmission coefficients from the aperture of a symmetric MIM waveguide (Ag-air-Ag) due to the fundamental TM mode over a range of aperture width $\text w$ at three different wavelengths: magnitudes at $\lambda_0=600$ nm (a), $\lambda_0=900$ nm (b), $\lambda_0=1500$ nm (c), and  phases at $\lambda_0=600$ nm (d), $\lambda_0=900$ nm (e), $\lambda_0=1500$ nm (f).}
	\label{fig:MIM_symmetric_3wavs}
\end{figure}

Symmetric finite MIM waveguides have been studied for the reflection and transmission of the SPP modes using full-wave analysis methods as well as by approximate models with closed form expressions on the reflection and transmission coefficients. Since the main contribution here is to propose a new and improved approximate model, based on the surface impedance boundary condition at the plane of the aperture, the starting point and the benchmark for the symmetric case have inevitably been the work by Lalanne et al.'s~\cite{Lalanne2005theory,Lalanne2006approximate}. Hence, the symmetric structure with silver-air-silver composition was chosen to validate the proposed model and to compare against the full-wave FDTD method and Lalanne et al.'s approximate model, as shown in Fig.~\ref{fig:MIM_symmetric_3wavs}, for a wide range of insulator widths ($0.01 <w/ \lambda_0<1.5$) at three different free-space wavelengths,  600~nm, 900~nm and 1500~nm. To furtheremphasize the contributions of the higher-order modes and the convergence of the proposed method, the results due to the increasing number of modes in the MIM waveguide ($\text{TM}_0$ for 1 mode, $\text{TM}_0$ and $\text{TM}_2$ for 2 modes, and $\text{TM}_0$, $\text{TM}_2$ and $\text{TM}_4$ for 3 modes) are given separately in the figure. Note that, owing to the symmetry of the structure, only the even order modes are used and that it seems only the first two would be sufficient for a reasonable accuracy for this example.

There are few observations that one can draw from the results presented in~Fig.~\ref{fig:MIM_symmetric_3wavs}, which are summarized as follows: i. The single mode case with the proposed method (dark blue line) performs well for a narrow slit ($w / \lambda_0<0.5$) but fails beyond. This is expected as the single mode would be the only significant mode when the slit is narrow. However, when the width increases, the second mode whose field profile is shown in green in Fig.~\ref{fig:MIM_finite} starts contributing to the field profile in the slit significantly, hence the lack of its contribution causes the substantial deviation. Also it appears that the model proposed in~\cite{Lalanne2006approximate} (red dashed line) performs better even though it uses the fundamental mode only. ii. As stated, with the addition of the second mode (magenta dashed line), the proposed method performs extremely well, better than any approximate model available. iii. Adding the third mode further improves the results, though slightly, for the wider slits (solid, light blue), which highlights the convergence of the model with the increasing number of modes.  iv. The SPP transmittance dip moves towards the ideal cut-off width (when the walls are assumed to be PEC) for the $\text{TM}_2$ mode ($w/\lambda_0 = 1$) as the wavelength increases, which would be expected because of the increasing conductance of the silver and making the walls more like PEC. v. Perhaps the most important observation is that the proposed model captures the nature of the problem and follows almost the same trend as the reference results, both in magnitude and phase, for a wide range of slit widths and wavelengths, which is quite encouraging but should be validated on asymmetric structures as well. 

\subsection{Asymmetric MIM aperture}
\begin{figure}[h]
	\centering
	{\includegraphics[width=15cm]{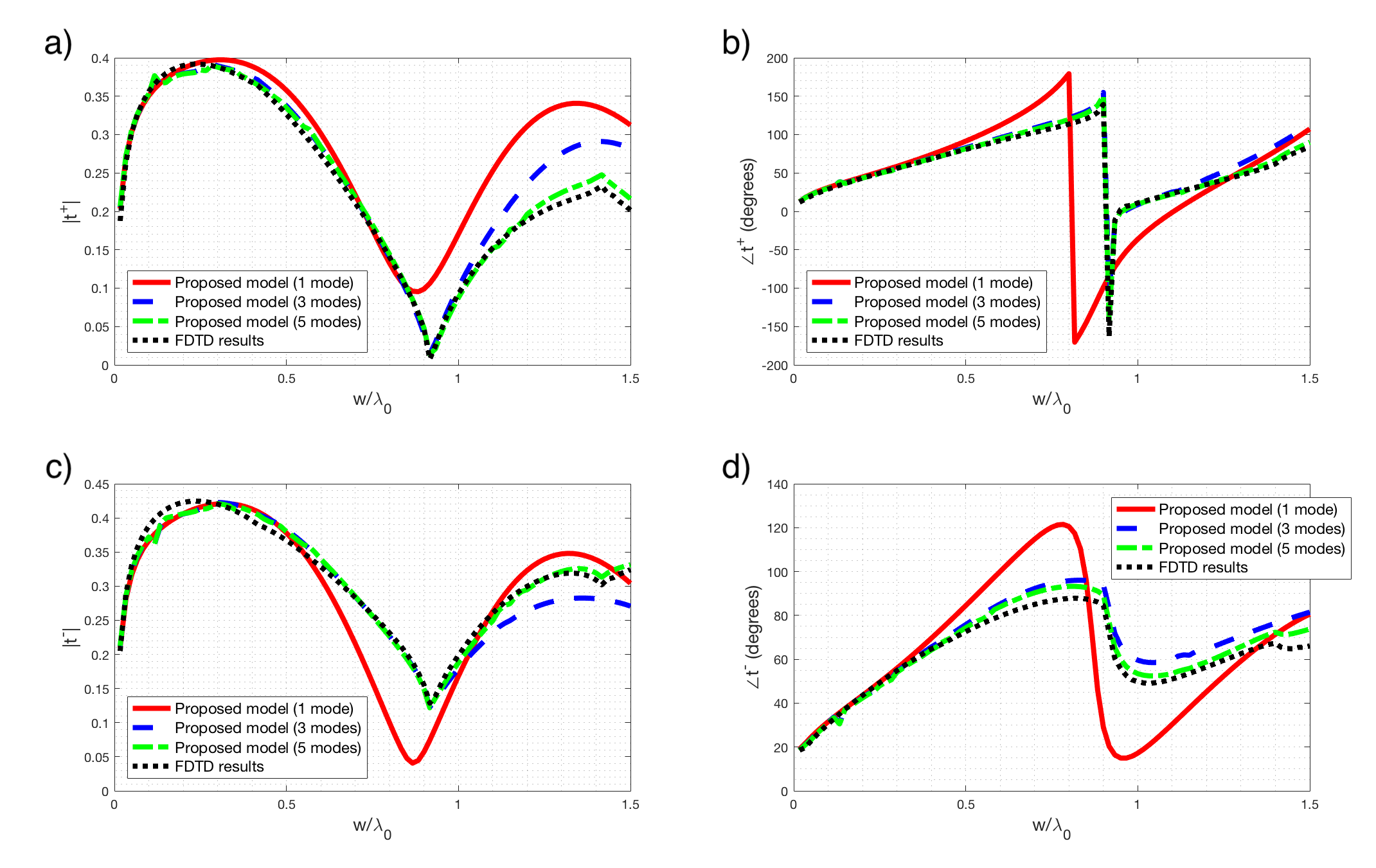}}
	\caption{Transmission coefficients from the aperture of a MIM waveguide (Ag-air-Cu) due to the fundamental TM mode at $\lambda_0=600$~nm  for a range of width of the aperture $w$: a) Transmission magnitude $|t^+|$ at $z=w/2$; b) Transmission phase $\angle t^+$ at $z=w/2$; c) Transmission magnitude $|t^-|$ at $z=-w/2$; b) Transmission phase $\angle t^-$ at $z=-w/2$.}
	\label{fig:MIM_600nm_Ag_Cu}
	\end{figure}

As stated above, asymmetry may be due to different metals used on both sides of the slit or different planar dielectric layers inside the slit or both. For the first example, SPP transmission from the aperture of a MIM waveguide with silver-air-copper configuration (silver for $z > w_1$ and copper for $z < -w_2$) is studied at the free-space wavelength of 600~nm. The choice of rather low operational wavelength is due to achieving a high degree of structural asymmetry and being able to assess the model under more general and unfavorable conditions. Note that, due to the asymmetry, the transmissions to both sides of the aperture are different and given separately in Fig.~\ref{fig:MIM_600nm_Ag_Cu}, where (a) and (b) give the magnitude and phase of the transmission to the upper side, respectively, while (c) and (d) show them on the lower side of the aperture. Contrary to the symmetric case, both odd and even modes need to be involved and in most cases the first five modes would be sufficient to approximate the transmissions with good accuracy for broad range of aperture widths and operation wavelength. As in the case of the symmetric MIM waveguide, the model reproduces the SPP transmission amplitudes and phases on both sides of the aperture with very high accuracy, as shown in Fig.~\ref{fig:MIM_600nm_Ag_Cu}. The accuracy and convergence of the model for the asymmetric geometry were crucial for the validation of the model, as it employs the impedance of one of the metals (the one on the side where the transmission is calculated; silver for $t^+$ and copper for $t^-$ in Fig.~\ref{fig:MIM_600nm_Ag_Cu}) as the impedance boundary for the entire left-hand region. In both symmetric and asymmetric examples, it has been observed that the transmission coefficients, more importantly their trends over the range of widths or wavelengths, can be predicted with improved accuracy as the number of modes in the slit increases, implying that the proposed model captures the nature of the problem.
	
\begin{figure}[h]
	\centering
	{\includegraphics[width=15cm]{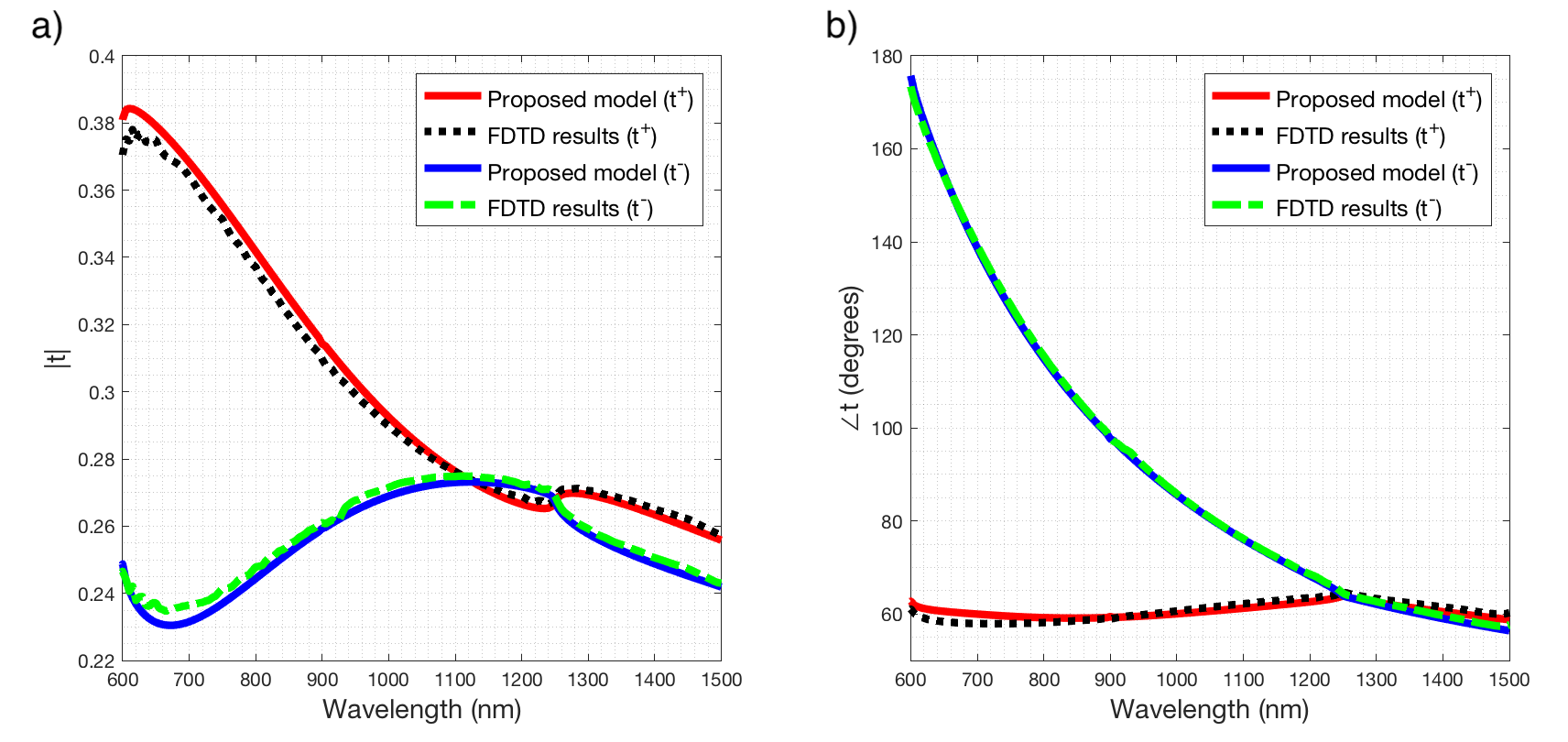}}
	\caption{Transmission coefficients from the aperture of a MIM waveguide (Ag-$\text{SiO}_2$-air-Ag, $w_1=300$~nm and $w_2=100$~nm) due to the fundamental TM mode for a range of wavelength: a) Transmission magnitudes $|t^+|$ and $|t^-|$ at $z=w_1$ and $z=-w_2$, respectively; b) Transmission phases $\angle t^+$ and $\angle t^-$ at $z=w_1$ and $z=-w_2$, respectively, as a function of wavelength between 600~nm and 1500~nm.}
	\label{fig:MIM_layer4_300_to_100}
\end{figure}

In the following example, asymmetry has been introduced in the insulator region by splitting it into two different dielectric materials, hence the whole structure has become a four-layered finite MIM waveguide with silver being the metal on both sides, and silica ($\varepsilon_{d1}=\varepsilon_{\text{SiO}_2}=2.1025$) and air as the layers of insulator region. The geometry is studied over the same range of free-space wavelength (600~nm-1500~nm) with the pre-defined widths of the dielectric layers as $w_1=300$~nm and $w_2=100$~nm. As for the previous asymmetric structure, the first five guided modes (up to $\text{TM}_4$) are included in the computation with excellent results, as shown in Fig.~\ref{fig:MIM_layer4_300_to_100}. It is worth repeating that, again the predictions of the transmission coefficients, both in magnitude and phase, capture every minute bends and twists of the transmissions. 

One interesting observation of the SPP transmission for this particular structure is the effective symmetry of the surface wave generation at both ends of the aperture for one particular wavelength of approximately 1250~nm. Despite the apparent asymmetry of the structure, the plasmonic responses of the configuration on the aperture side walls are approximately symmetric both in terms of amplitude and phase. This behavior is observed to be strictly dependent on the geometry, as well as the dielectric properties of the intermediate stacks, which deserves further study in the future. 

\begin{figure}[h]
	\centering
	{\includegraphics[width=15cm]{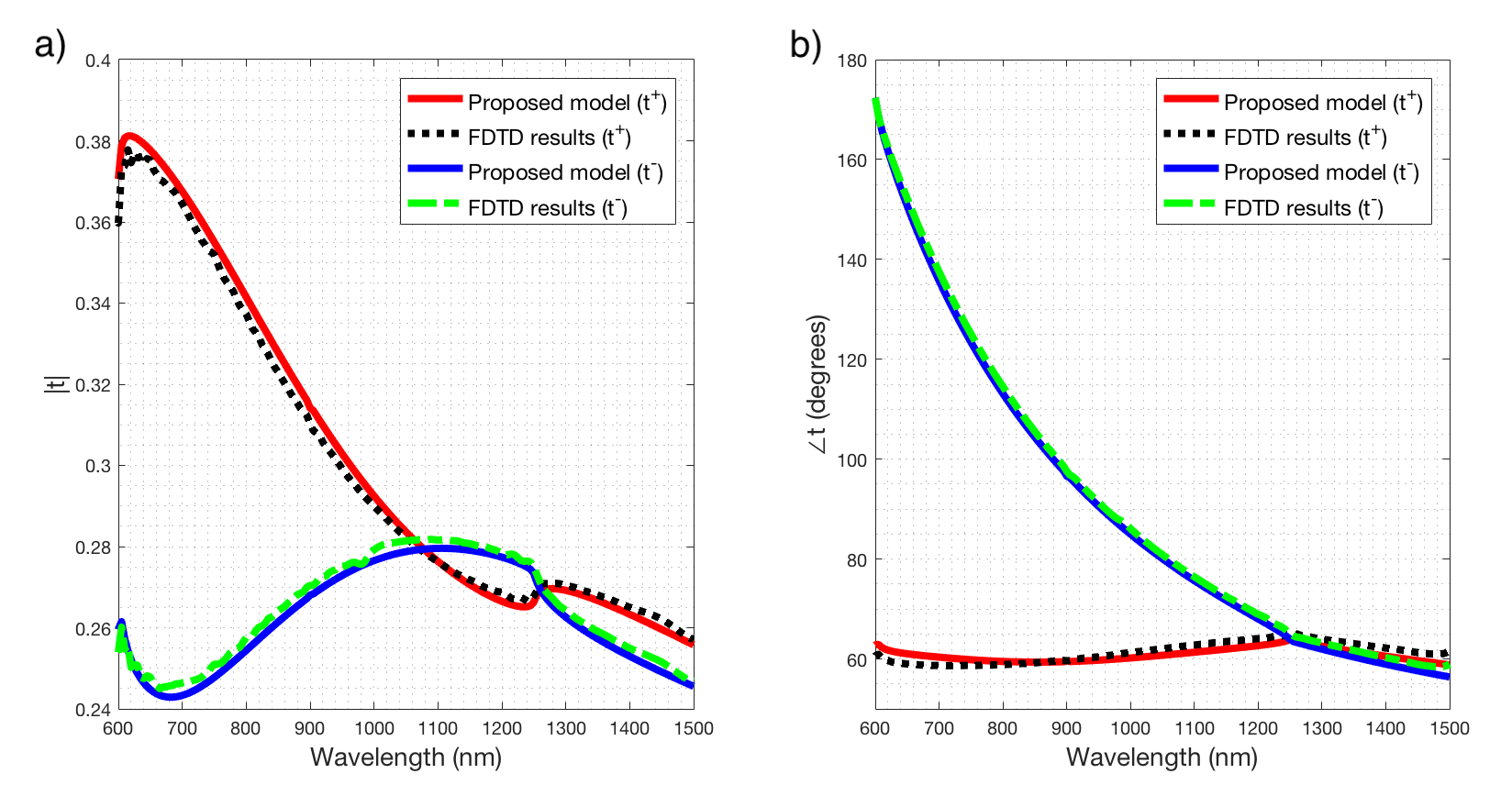}}
	\caption{Transmission coefficients from the aperture of a MIM waveguide (Ag-$\text{SiO}_2$-air-Cu, $w_1=300$~nm and $w_2=100$~nm) due to the fundamental TM mode for a range of wavelength: a) Transmission magnitudes $|t^+|$ and $|t^-|$ at $z=w_1$ and $z=-w_2$, respectively; b) Transmission phases $\angle t^+$ and $\angle t^-$ at $z=\text{w}_1$ and $z=-w_2$, respectively, as a function of wavelength between 600~nm and 1500~nm.}
	\label{fig:MIM_layer4_ag_cu_300_to_100}
\end{figure}

For the sake of completeness, all possible asymmetries in MIM waveguides are covered by introducing asymmetry in metals (using different materials on both sides) and asymmetry in the insulating region, by combining the geometries in the previous two examples. That is, the resulting geometry becomes a four-layered MIM configuration with the layers from top to bottom as silver, silica, air and copper. With the same operational parameters as in the previous example, the agreements between the predictions and the reference data are again excellent with a similar observation on the  symmetric response of the SPP at a particular wavelength, as shown in Fig~\ref{fig:MIM_layer4_ag_cu_300_to_100}.

\begin{figure}[h]
	\centering
	{\includegraphics[width=14cm]{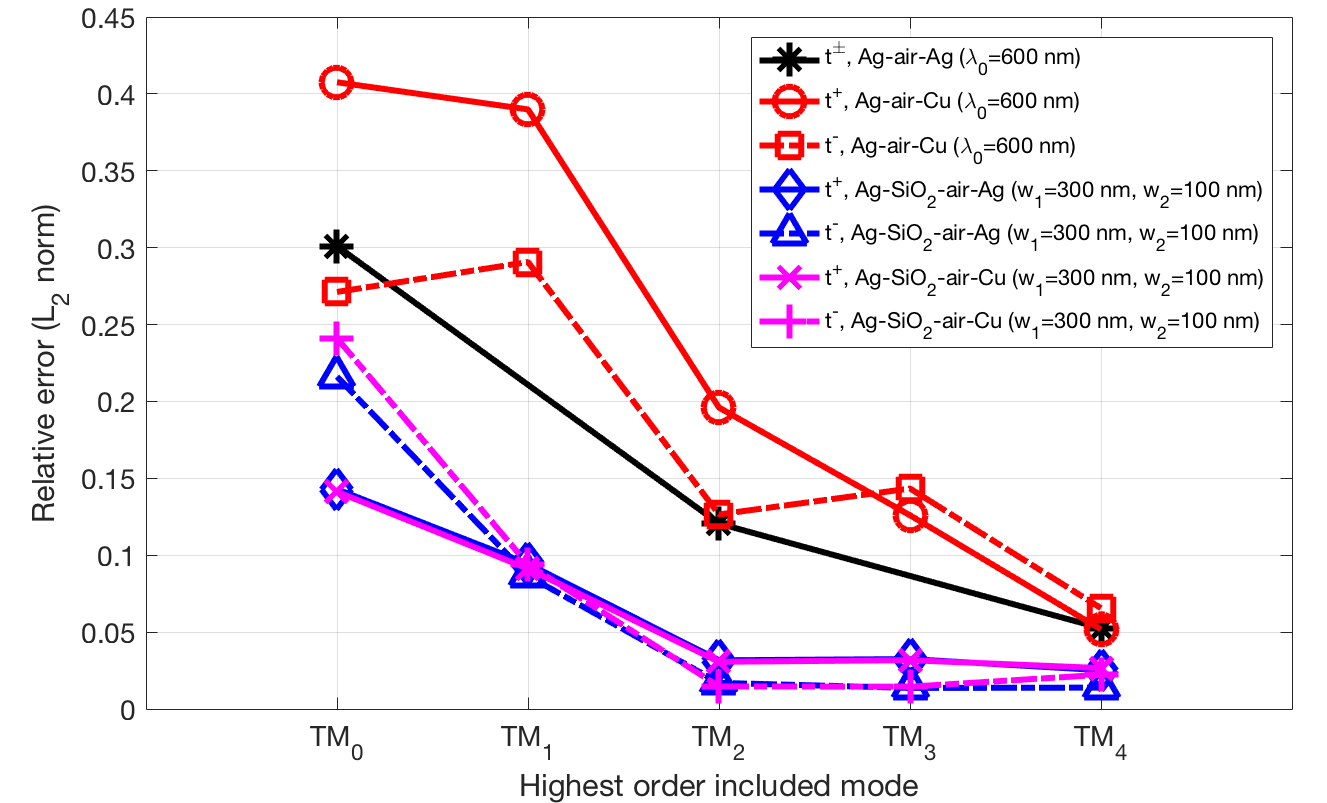}}
	\caption{Relative errors for the transmission coefficients presented in Figs.~\ref{fig:MIM_symmetric_3wavs}-\ref{fig:MIM_layer4_ag_cu_300_to_100} as a function of the number of modes used in the approximation.}
	\label{fig:model_fdtd_l2_error}
\end{figure}

After having completed the presentation of the cases, the relative errors of the approximations for all the data shown in Figs.~\ref{fig:MIM_symmetric_3wavs}-\ref{fig:MIM_layer4_ag_cu_300_to_100} have been calculated and given in Fig.~\ref{fig:model_fdtd_l2_error}. A few observations are in order: i. The model always converges as the number of modes increases. ii. However, the convergence is not uniform for all cases. iii. For the two curves that show slight increase and decrease in the error (red lines), both are for the asymmetric metal case, Fig.~\ref{fig:MIM_600nm_Ag_Cu}, which could be considered approximately symmetric as the SPP field profile is dominantly symmetric. As a result, using the odd modes as the basis and testing functions for the MoM application causes either a small improvement or small additional error. Judicious choice would be to use only the even modes as the basis functions, as was done for the symmetric case (black line). iv. The first five modes, at most, have been sufficient to achieve the desired error criterion for almost all cases that have been studied so far.  

As a final note, despite its wide-range of applicability and robustness, the proposed model is also an approximate model and has a limitation at shorter wavelengths due to the use of  the zeroth-order surface-impedance approximation in (\ref{equivalent-current-aperture}).  That is, the equivalence principle is violated for metallic layers at shorter wavelengths because its surface impedance ($Z_s = Z_0 /\sqrt{\epsilon_m}$) can no longer represent the true impedance of the material used in the definition of the equivalent magnetic current density in (\ref{equivalent-current-aperture}). The remedy for this is to use a higher order approximation for the effective impedance of the metallic layer for the equivalence principle to hold true~\cite{TretyakovAnalytical,Gholipour2013}, at the expense of a considerable computational cost. This is the reason why all the data presented here starts at the free-space wavelength of 600~nm and goes well beyond 1500~nm, although the model can predict the transmissions in some cases down to 500~nm of free-space wavelength.

\section{Conclusion}
A model based on the modified equivalence principle has been developed to predict the SPP transmission from the aperture of a finite MIM waveguide to the sidewalls of the aperture with good accuracy and computational efficiency. It has been shown that the model is robust and quite successful for a wide range of wavelengths and MIM configurations, as validated by comparing the results with those obtained from a full-wave em solver using the FDTD method. The convergence of the model with the number of modes employed in the formulation has also been discussed, and shown that even for a complex MIM configuration, it converges with a small number of modes and is quite robust for the variations of the geometric parameters. The limitations of the model have also been discussed with possible remedies, though computationally expensive, and it is recommended that as long as the model is used for the free-space wavelength of 600~nm and above, it captures the trend of the transmission well and always converges.

\bibliographystyle{apsrev4-1} 
\bibliography{spp_paper}

\end{document}